# Heterogeneous field response of hierarchical polar laminates in relaxor ferroelectrics

Hao Zheng[1, *], Tao Zhou[2], Dina Sheyfer[3], Jieun Kim[4], Jiyeob Kim[4], Travis D. Frazer[1, #], Zhonghou Cai[3], Martin V. Holt[2], Zhan Zhang[3], J. F. Mitchell[1], Lane W. Martin[4, 5, *, †], and Yue Cao[1, *]

[1] Materials Science Division, Argonne National Laboratory, Lemont, IL 60439, USA.
[2] Center for Nanoscale Materials, Argonne National Laboratory, Lemont, IL 60439, USA.
[3] X-ray Science Division, Argonne National Laboratory, Lemont, IL 60439, USA.
[4] Department of Materials Science and Engineering, University of California Berkeley, Berkeley, CA 94720, USA.
[5] Materials Sciences Division, Lawrence Berkeley National Laboratory, Berkeley, CA 94720, USA.

\#   Present address: KMLabs, Boulder, CO 80301, USA
†   Present address: Department of Materials Science and NanoEngineering and the Rice Advanced Materials Institute, Rice University, Houston, TX 77005, USA
* For further information: haozheng@anl.gov (H.Z.), lwmartin@rice.edu (L.W.M.), yue.cao@anl.gov (Y.C.)

## Abstract

Relaxor ferroelectrics are a class of materials that are widely perceived as deriving their exotic properties from structural heterogeneities. Understanding the microscopic origin of the superior electromechanical response requires knowledge not only concerning the formation of polar nanodomains (PNDs) built from individual atoms but more importantly the spatial distribution of PNDs over longer distances. The mesoscale PND arrangement is shaped by the interactions between these domains and, in turn, dictates the electric-field driven PND response directly relevant to the macroscopic material properties. Despite decades of research, current understanding of relaxors at this important mesoscale regime remains limited, hindered by structural complexities that spans multiple orders of magnitude of length scales. Here, we show the emergence of mesoscale lattice order that we name "polar laminates" in the canonical relaxor ferroelectric $0.68PbMg_{1/3}Nb_{2/3}O_3$-$0.32PbTiO_3$ (PMN-0.32PT) using X-ray coherent nano-diffraction. These laminates are nematic with a size of ~350 nm and arise from the staggered arrangement of ~13 nm monoclinic PNDs along the <110> of the pseudocubic lattice. The spatial distribution of *c*-axis strain is directly correlated with the tilting of the PNDs and is most prominent between the laminates. Further *operando* nano-diffraction studies demonstrate heterogeneous electric-field-driven responses. The most active regions tend to reside inside the laminates while the spatial pinning centres are between the laminates. This observation reveals the hierarchical assembly of lattice order as a novel form of electron and lattice self-organization in heterogenous materials and establishes the role of such mesoscale spatial arrangement in connecting the nanoscale heterogeneity and macroscopic material properties. These findings provide a guiding principle for the design and optimization of future relaxors and may shed light on the existence of similar behaviour in a wide range of quantum and functional materials.

**Main text**

The properties of crystalline materials and their responses under the external stimuli are determined by their structures across the relevant length scales. Classically, the types of responses a material manifests are perceived as being defined by the combination of atomic- (*i.e.*, at the 0.01-1 nm length scale) and micro-structures (*e.g.*, grain structure at the 1,000-100,000 nm length scale). In a wide range of materials, however, additional complexity can arise in between from mesoscopic interactions, thus bridging across those length scales traditionally considered. For example, existing studies observed ubiquitous spatial heterogeneities across a wide range of quantum materials, where nanoscale lattice deformations deviate from the average atomic structure[1–5]. It remains largely elusive, however, as to how these spatial heterogeneities affect local and macroscopic material properties in response to external stimuli. This missing link is important for fully understanding the structure-property relationship and calls for spatial-resolved measurements under *operando* conditions.

An exemplary class of materials strongly influenced by spatial structural heterogeneity are the relaxor ferroelectrics, where heterogeneity is widely believed as being pivotal to the manifestation of exceptional electromechanical properties. In contrast to regular ferroelectrics, relaxor ferroelectrics feature almost non-saturating strain response under the application of several orders of external electric fields[6–10]. The relaxor behaviour is most prominent near the morphotropic phase boundary (MPB) where the lattice should be considered as a solid solution of different space-group symmetries[1,11–13]. X-ray-[14–22] and neutron-based[20,23–30] scattering experiments have revealed that not only does the unit-cell structure evolve in the form of charge polarization rotation and space group symmetry change, but that this atomic-structural change is intimately connected to the local polar order on the scale of 5-10 nm, corresponding to the formation of polar nanodomains (PNDs)[31–38]. Current efforts to understand relaxor behaviour have focused predominantly on the interplay between the atomic structure and the PNDs[14–38]. In contrast, little attention has been given to longer-range interactions and length scales, particularly on the mesoscale between 10-1,000 nm[39–42]. While PMN is widely acknowledged as being spatially heterogeneous, the possible role of interactions among the PND has not been demonstrated. In canonical ferroelectrics with micron-sized domains, inter-domain interactions and the free energy of the domain walls are indispensable to the stabilization of the ferroelectric order and the macroscopic electric-field response. This begs the question of how the PNDs are spatially arranged over larger length scales and how such arrangement could enhance the electromechanical response under electric field in actual devices.

Here, using *operando* coherent nano-diffraction (CND)[43], we show the formation of what we call "polar laminates" that comprise of unidirectional staggered monoclinic PNDs in the canonical relaxor ferroelectric $(1-x)$PbMg$_{2/3}$Nb$_{1/3}$O$_3$-$(x)$PbTiO$_3$ (PMN-$x$PT). The field-driven responses are drastically different within and between these laminates. For this work, PMN-0.32PT was chosen because of its proximity to the MPB, and the PMN-0.32PT thin films were synthesized on SmScO$_3$ (SSO) (1 1 0)$_o$ substrates (where the "o" denotes orthorhombic indices) with a 25 nm Ba$_{0.5}$Sr$_{0.5}$RuO$_3$ (BSRO) bottom electrode[44]. A second 25 nm BSRO layer was grown *in situ* on top of the PMN-0.32PT and fabricated into circular capacitor structures to serve as the top electrode for electric-field-driven studies (Figure 1a). Details of the material synthesis and the characterization of the PMN-0.32PT devices are provided (**Methods** and **Extended Dataset 1**). The average lattice structure of the SRO/PMN-0.32PT/BSRO/SSO (1 1 0)$_o$ heterostructures was obtained from reciprocal space mapping (RSM) studies about the 002$_{pc}$-Bragg peak (where "pc" denotes pseudocubic indices) acquired using parallel X-ray beams (Figures 1b, c). Details of the X-ray characterization are provided (**Methods**). The Miller indices H, K, and L are labelled using the pseudocubic lattice parameters of the SSO substrate with $c_{SSO,pc} = 3.992$ Å. Hereafter, unless otherwise noted, the lattice parameters are those of the PMN-0.32PT. The 002$_{pc}$-Bragg peak of the PMN-0.32PT is found at L = 1.95, corresponding to an average lattice parameter $c_{pc} = 4.098$ Å, slightly longer than that of the bulk crystal[45]. This elongation comes from the slight compressive strain in plane on the PMN-0.32PT due to epitaxy with the SSO substrate. Despite a minor in-plane anisotropy ($a_{pc} = 3.991$ Å and $b_{pc} = 3.983$ Å), the diffuse scattering of the PMN-0.32PT about the 002$_{pc}$-Bragg peak is essentially isotropic in the film plane (Figure 1c). The line profiles of the diffuse-scattering pattern along H and K are both Lorentzians with a full-width-at-half-maximum (FWHM) of 0.0075 Å$^{-1}$ (Figure 1d), corresponding to a correlation length of ~13.4 nm which can be taken as the characteristic size of the PNDs in the film plane. Further analysis of the diffuse-scattering pattern revealed that the average-atomic structure of the PMN-0.32PT in the heterostructure is monoclinic ($M_c$), consistent with prior reports[44]. On the length scale of the device (50 μm in diameter), the PNDs are twinned with finite lattice tilts along either [1 0 0]$_{pc}$ or [0 1 0]$_{pc}$ deviating from 90° (Figure 2a-c), which we name $\Delta\alpha$ or $\Delta\beta$, respectively.

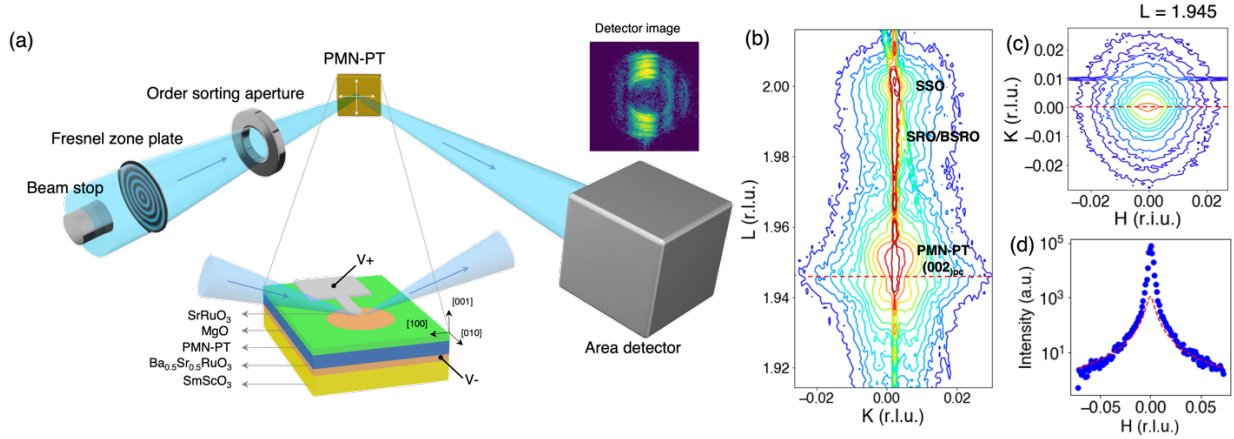

**Figure 1: The PMN-0.32PT thin film device and the experimental setup of the *in-operando* X-ray coherent nano diffraction (CND). a** Sketch of the experimental setup of in-situ X-ray CND. The electric field is applied along the [001]$_{pc}$ direction. Raster scans using the nanobeam reveal the local domain structure in PMN-0.32PT. Bottom inset: the BSRO/PMN-0.32PT/BSRO/SSO heterostructure. Top inset: A typical nano diffraction pattern from the PMN-0.32PT (002)$_{pc}$ Bragg peak. **b, c** H-cut and L-cut from the reciprocal space map of the PMN-0.32PT/BSRO/SSO heterostructure. The miller indices H, K, and L are labelled using the lattice parameters of the SSO substrate. The (002)$_{pc}$ peak of PMN-0.32PT shows an almost circular diffuse scattering pattern. **d** Diffuse scattering intensity extracted from the dashed lines in **c**, and red lines show the Lorentz fitting to the profile of the diffuse scattering, i.e., $I = \frac{I_0}{\pi}\frac{\Gamma}{(q-q_0)^2+\Gamma^2}$. A fitted value of Γ=0.00755Å$^{-1}$ is obtained corresponding to a correlation length of 132 Å.

A schematic of the *operando* CND experiment is shown (Figure 1a) wherein the PMN-0.32PT-based capacitor devices were pre-aligned and mounted in a reflection geometry with the [0 0 1]$_{pc}$ and [1 0 0]$_{pc}$ in the scattering plane (**Methods**). Upstream of the sample, highly coherent 10 keV X-ray photons were focused to a diameter of 25 nm (FWHM) using a Fresnel zone plate, with a convergence angle of 0.5°. The FWHM of the focused X-ray is comparable to the size of the PNDs, allowing access to the spatial arrangement of and interactions among PNDs. The sample theta and detector two-theta angles were tuned to access the 002$_{pc}$-Bragg reflection of the PMN-0.32PT film. Due to the difference in the *c* lattice constants under the pseudocubic notation, the nano-diffraction intensity comes predominately from the PMN-0.32PT and not from the SSO or BSRO. A typical CND pattern (inset, Figure 1a) encodes information about the local lattice deformation within the focused X-ray photon footprint. Analysing the distribution of diffraction intensity allows for quantification of the local lattice strain $\epsilon_c$ and tilts (*i.e.*, $\Delta\alpha$ and $\Delta\beta$). Details of the simulation and the extraction of lattice distortion parameters are provided (**Methods**, **Extended Datasets 4, 5, and 6**).

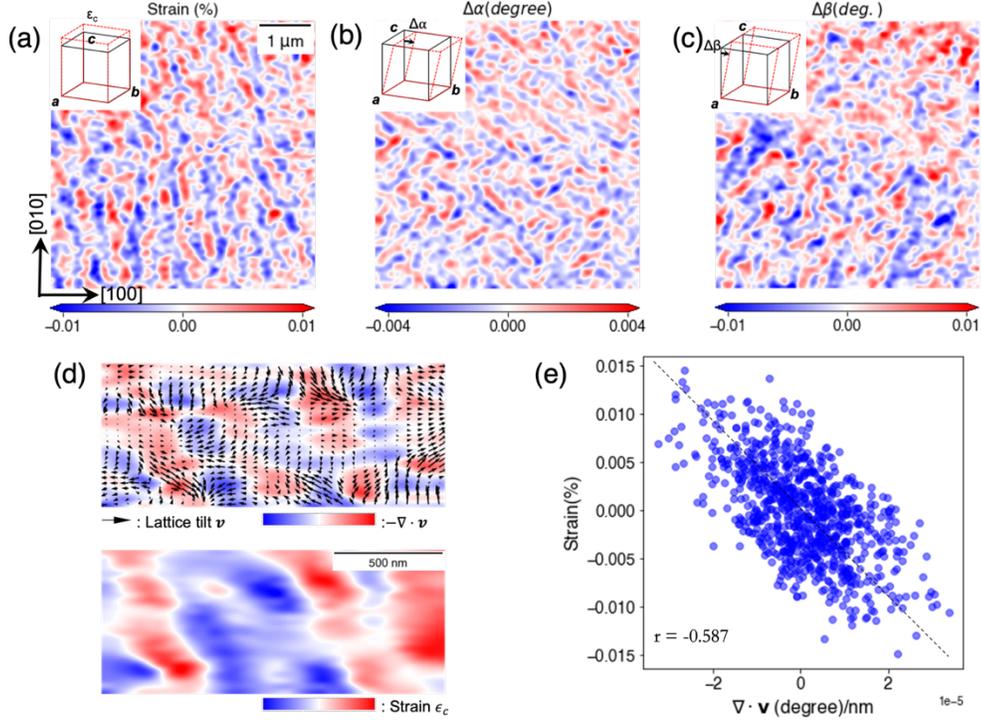

**Figure 2: Mesoscale distributions of the lattice distortions. a-c** The distortion of the c lattice vector away from the average pseudocubic unit cell can be decomposed into three components: strain $\varepsilon_c$, lattice tilt $\Delta\alpha$, and lattice tilt $\Delta\beta$. Spatial distributions of $\varepsilon_c$ (**a**), $\Delta\alpha$ (**b**), and $\Delta\beta$ (**c**) were obtained from the CND over a 5 µm × 5 µm area. Schematics of these distortions are displayed in the insets on the top left corner of each map. **d** Correlation between strain and the divergence of the projected lattice tilts. Upper panel: projected lattice tilts (arrow) and the calculated divergence (colour map). Lower panel: the distribution of measured strain. **e** Scatter plot of strain and divergence of lattice tilt. A Pearson correlation coefficient r = -0.587 confirms the correlation the two quantities.

Typical distributions of $\epsilon_c$, $\Delta\alpha$, and $\Delta\beta$ extracted from the raster scan of a 5 µm × 5 µm field of view are shown (Figure 2a-c). The spatial distributions of these parameters exhibit three features that are distinct from those in standard ferroelectrics. First, the spatial distribution of the *c*-axis strain $\epsilon_c$ is not directly correlated with that of either $\Delta\alpha$ or $\Delta\beta$, in sharp contrast to canonical ferroelectrics (*e.g.*, BiFeO$_3$[46]). The lack of direct correlation arises from the small size of the PNDs, and fundamentally reflects the nature of PMN-0.32PT as a solid-state solution near the MPB. Second, a spatial correlation does exist between the *c*-axis strain and the *relative* change of the projected lattice tilt $\boldsymbol{v} = (\Delta\beta, \Delta\alpha)$. To visualize this, a zoomed-in view of the lattice distortion (2.5 µm ×1 µm) is displayed (upper panel, Figure 2d). The colour map shows the distribution of the divergence $-\nabla \cdot \boldsymbol{v}$ while the arrows represent the projected lattice tilts $\boldsymbol{v}$. The prefactor of -1 was multiplied to the divergence for direct visual comparison with the distribution of $\epsilon_c$ over the same area (lower panel, Figure 2d). Indeed, a Pearson coefficient *r* = -0.587 can be obtained from the correlation plot (Figure 2e). The correlation between $\epsilon_c$ and $\nabla \cdot \boldsymbol{v}$ is nontrivial and highlights the importance of the domain walls between the PNDs. Specifically, local maxima and minima of $\nabla \cdot \boldsymbol{v}$ correspond to the regions where the neighbouring $\boldsymbol{v}$ tilts towards opposite (180°) directions. These regions contain domain walls between the twinned monoclinic PNDs. Third, the PNDs self-organize on the mesoscale and form a nematic order almost parallel to the <1 1 0>$_{pc}$. The nematic order is unidirectional, which is best seen from the stripy distribution of $\Delta\alpha$ (Figure 2b) and of $\Delta\beta$ (Figure 2c), albeit less clearly in the latter.

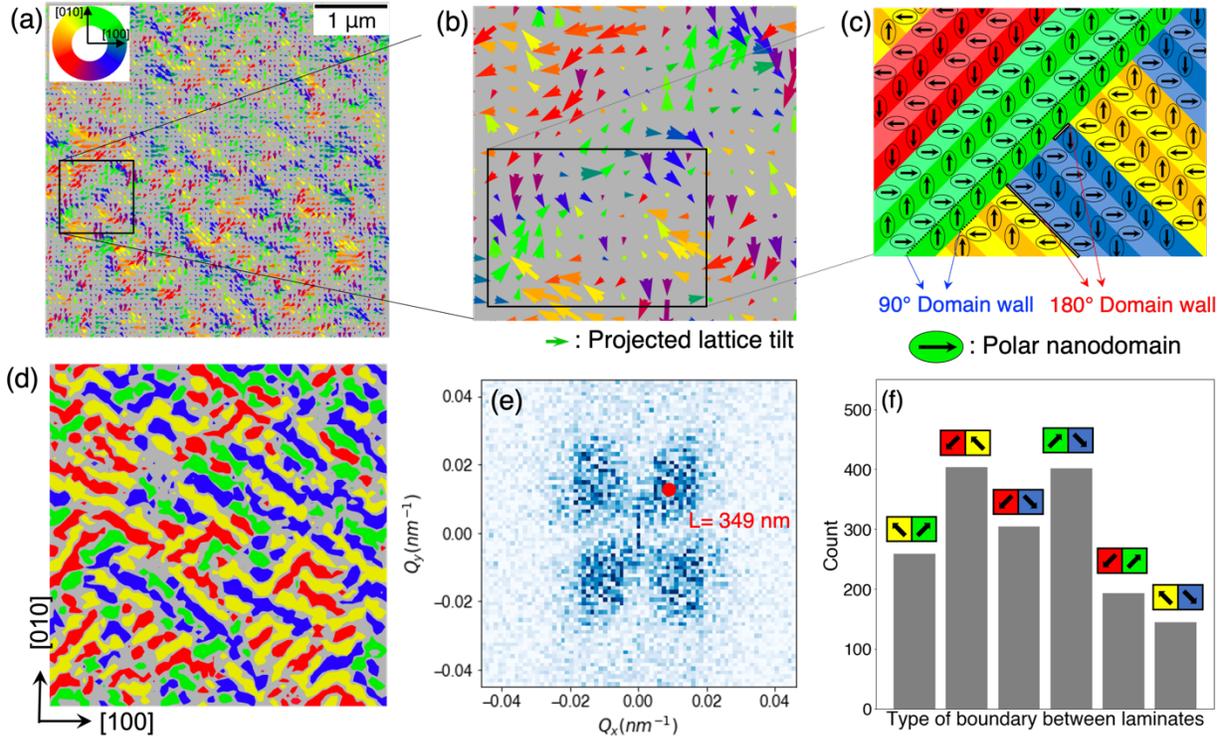

**Figure 3: Nematic polar laminates and its relation to PNDs. a** Vector field representing the spatial distribution of projected lattice tilt from the same area shown in Figure 2b The orientations of lattice tilt are labelled with colours indicated in the colour wheel. **b** A zoomed-in view showing the spatial arrangement of lattice tilts. A polar laminate consists of neighbouring sites with similar lattice tilts and tends to align along the diagonal $<110>_{pc}$ directions. **c** Cartoon schematic of polar laminates. Inside each laminate, staggered $M_c$ domains with perpendicular polarizations (black arrows) give rise to the 90° PND walls and extend diagonally. In contrast, boundaries between laminates host a mixture of 90° and 180° PND walls. **d** The classification of the polar laminate structure is generated by clustering the lattice tilt with similar orientations. The preferred orientation of the lattice tilt inside each laminate is represented by its colour. **e** Fast Fourier transform of the spatial map shown in **a**. The butterfly pattern reveals a four-fold diagonal arrangement of polar laminates with a correlation length of 349 nm. **f** Statistics of the boundaries between laminates. There are substantially more boundaries between orthogonal laminates than between antiparallel ones.

Further statistical analysis of the projected lattice tilts $v$ revealed that the nematic order originates from the self-organization of the PNDs. The distribution of $v$ from the same field of view (as in Figure 2b) is shown (Figure 3a). Lattice tilts with similar orientations are labelled with like colours and segregate into diagonal patches consistent with our observations of nematic order. Interestingly, a large fraction of the lattice tilts point away from the $<1\,0\,0>_{pc}$, and instead lie along the $<1\,1\,0>_{pc}$ (Figure 3b). Note, the PNDs in PMN-0.32PT are monoclinic $(M_c)^{22}$, thus, the diagonal lattice tilts should be interpreted as the existence of PNDs sharing 90° head-to-head or tail-to-tail domain walls within the X-ray photon footprint (Figure 3c), with the total charge polarization and lattice tilt along the $<1\,1\,0>_{pc}$. This interpretation is because the focused X-ray beam contains several PNDs within the illuminated volume. Specifically, the incidence angle of the X-ray is 17.5° with an in-plane footprint of 25 nm × 80 nm; ten-times larger than the size of the PNDs.

We refer to the diagonal arrangement of 90° PNDs over a continuous region that correspond to the nematic order as "polar laminates" (Figure 3d). The proliferation of these 90° domain walls inside the polar laminates might be attributed to the smaller amount of charge accumulated on these domain walls compared to alternative domain arrangements, hence making them lower in energy. Interestingly, similar diagonal arrangements correlated with a 90° phase shift were also observed on bulk PMN-0.40PT single crystals using piezoresponse force microscopy[47], and are indicative of the 90° intra-laminate domain walls. The observation here provides direct evidence of the polar-laminate structure, thanks to the much longer X-ray penetration depth beyond the top surface of the material.

We classify polar laminates into four groups using the experimentally observed lattice tilts and identify their characteristic ordering vector and length scale. A classification of polar laminates can be generated by clustering neighbouring sites with similar lattice tilts (Figure 3d). The predominant orientation of lattice tilt inside the laminate is labelled using different colours (**Extended Dataset 7**). Here, green, yellow, red, and blue colours represent preferred lattice tilt along the $[1\ 1\ 0]_{pc}$, $[\bar{1}\ 1\ 0]_{pc}$, $[\bar{1}\ \bar{1}\ 0]_{pc}$, and $[1\ \bar{1}\ 0]_{pc}$, respectively. The two-dimensional fast Fourier transform (FFT) of the laminate distribution has a characteristic ordering wavevector centred at $Q = (0.012, 0.014, 0)$ nm$^{-1}$ (Figure 3e), corresponding to a length scale of ~350 nm in the sample plane characteristic of the polar laminates. The large length scale, well over 100 nm, prevented these polar laminates from being directly resolved in the RSM due to the limited momentum resolution. The direction of Q deviates slightly from the $[1\ 1\ 0]_{pc}$ possibly because of the minor difference between the lattice constants $a_{pc}$ and $b_{pc}$ in plane.

Following the classification, six types of inter-laminate boundaries have been identified. A histogram of these boundaries from the field of view used previously (Figure 3a) is provided (Figure 3f). There are several observations to note. First, boundaries between orthogonal laminates are preferred compared to those between antiparallel laminates. This orientational preference comes from the nature of the inter-laminate domain walls. Specifically, while the intra-laminate PND walls are entirely 90° head-to-head, the orthogonal inter-laminate boundaries consist of alternating 90° and 180° PND walls (Figure 3c). As the 180° domain walls are generally considered as higher in energy than the 90° walls[48], the boundaries between laminates are expected to be "harder" and less likely to move than the intra-laminate PND walls under an electric field (to be discussed later). Second, the inter-laminate boundaries can be categorized into two groups with 180° PND walls along the $[1\ 0\ 0]_{pc}$ (yellow-green and red-blue boundaries) and $[0\ 1\ 0]_{pc}$ (red-yellow and blue-green boundaries), respectively. The group with PND walls along the $[1\ 0\ 0]_{pc}$ have a slightly higher population, possibly due to the in-plane anisotropy from the SSO substrate epitaxy.

The discovery of polar laminates reveals a new kind of hierarchical spatial heterogeneity in relaxors and provides key insights into the interactions between PNDs. Specifically, individual atoms packed in ~ 0.4 nm unit cells form ~13 nm monoclinic PNDs, which subsequently assemble into polar laminates ~350 nm in size. Most intriguingly, these polar laminates are further organized in an almost self-similar manner, which is evidenced by the FFT (Figure 3e) reminiscent of the diffuse scattering in PMN-0.30PT[21]. This observation amounts to direct evidence that the PNDs and the polar laminates are hierarchically ordered over many magnitudes of lengths scales. Moreover, the formation of polar laminates highlights the cooperative interactions between PNDs. The dominance of 90° PND walls sharply contrasts that in typical ferroelectric materials and lowers the contribution of domain walls to the total free energy in relaxors due to the small domain size and the large number of domain walls.

The discovery of polar laminates provides the basis for revealing and understanding the spatially heterogenous field response of PMN-0.32PT. To do this, we performed *operando* CND under a DC electric field sequence of the form: $0$ kV·cm$^{-1}$ → $+540$ kV·cm$^{-1}$ → $-540$ kV·cm$^{-1}$ → $0$ kV·cm$^{-1}$. At each applied field, raster scans were performed to track the nanoscale evolution of lattice distortions (**Methods** and **Extended Dataset 8**). The distributions of $\epsilon_c$ and $\boldsymbol{v} = (\Delta\beta, \Delta\alpha)$ from a 2.5 $\mu$m × 2.5 $\mu$m field of view are displayed (Figure 4a1, 4b1). The relative strain distribution in the film plane generally stays intact under different electric fields (**Extended Dataset 8**), indicating strong spatial pinning of the PNDs. Such an effect is also evidenced by the evolution of the average strain along the field cycle. The strain averaged over the entire field of view exhibits a butterfly shape (Figure 4c) consistent with macroscopic property measurements. The butterfly loop is asymmetric relative to the field direction, with a more pronounced strain response under the negative bias. Such asymmetry was also observed in the polarization loop (**Extended Dataset 2**) as well as from other PMN-0.32PT devices we measured, possibly due to the asymmetric layouts of the top and bottom electrodes.

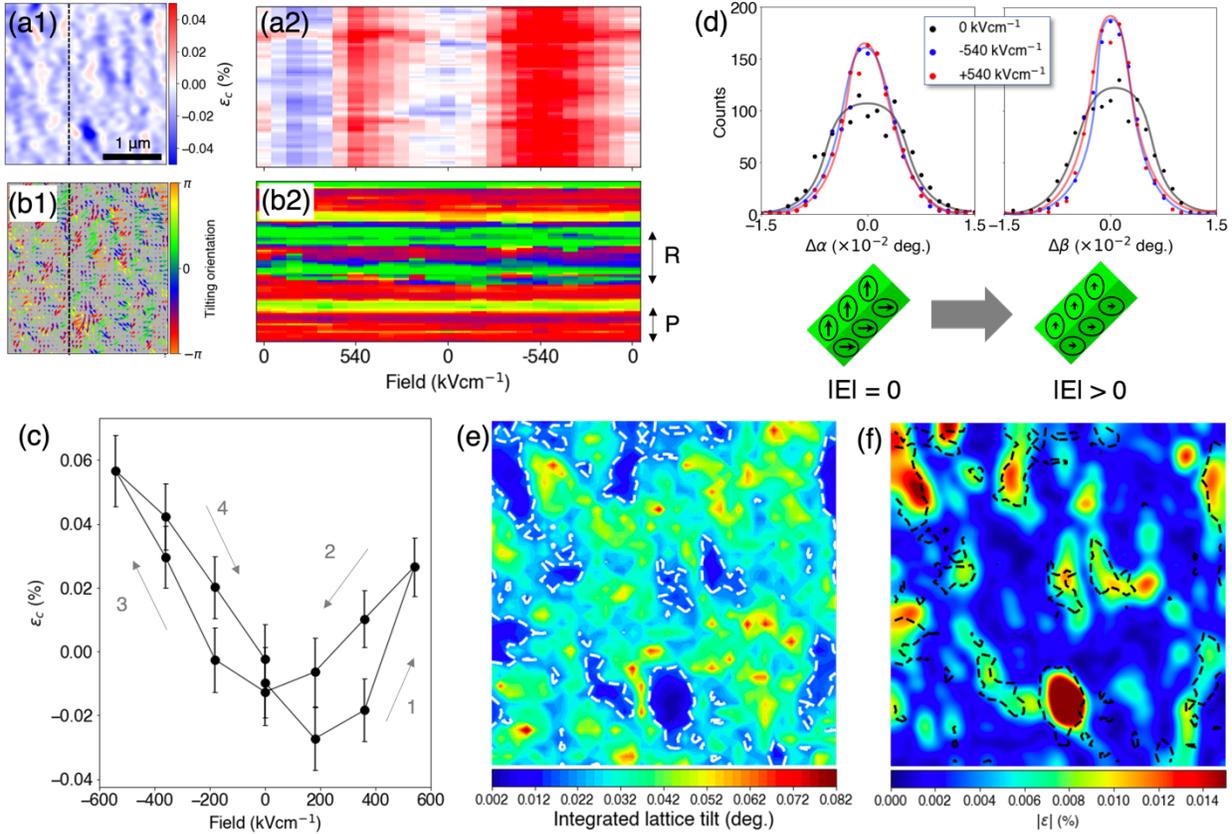

**Figure 4: Electric-field driven evolution of the polar laminates. a1** Strain distribution derived from in-situ CND scans on a 2.5 $\mu$m × 2.5 $\mu$m area at E = 0 kVcm$^{-1}$. **a2** Waterfall plot shows the voltage dependence of strain in a column of pixels (dashed line in **a1**) during a cycle of field 0 kVcm$^{-1}$ → +540 kVcm$^{-1}$ → 0 kVcm$^{-1}$ → -540 kVcm$^{-1}$ → 0 kVcm$^{-1}$ applied along the [001]$_{pc}$ direction. **b1** Map of projected lattice tilts from the same area. The orientation of projected lattice tilt is labelled with colours. The tilt angle is zero when the projected tilt is parallel to the [100]$_{pc}$ direction. **b2** Evolution of the orientation of projected lattice tilts from the same area. "P" and "R" correspond to the pinned and responsive regions, respectively. **c** Spatially averaged strain as a function of the applied electric field generated from the same area. The standard deviation of strain at each field is indicated by the error bar. **d** Statistics of lattice tilts $\Delta\alpha$ (upper left) and $\Delta\beta$ (upper right) at 0 kVcm$^{-1}$ and ±540 kVcm$^{-1}$. The electric field along the [001]$_{pc}$ direction induces a decrease of lattice tilts projected onto the sample plane, corresponding to a transition from the monoclinic to the orthorhombic lattice structure in PNDs (lower row). **e** The cumulative change of the projected lattice tilts. Red and blue colours represent the most and the least responsive regions. Dashed contours indicate the distribution of an intermediate integrated lattice tilt of 0.03 degree. **f** Spatial distribution of the absolute value of strain. An anti-correlation can be observed between strain and the response of the lattice tilts.

We further survey the evolution of the lattice tilts over the field cycle. On average, the electric field drives the lattice from the monoclinic $M_c$ phase towards the orthorhombic phase, consistent with the field-driven evolution of the average atomic structure[22]. Histograms of the projected lattice tilts are also provided (upper panels, Figure 4d). With increased electric field, the average lattice tilts shift towards the centre where $\Delta\alpha = \Delta\beta = 0$ (*i.e.*, an orthorhombic phase). Detailed inspection of the lattice-tilt evolution during the field cycle uncovers the highly heterogeneous field response. One-dimensional cuts of strain and projected lattice tilts were taken from the data (black lines, Figure 4a1, 4b1) and they are stacked as a function of the field cycle into "waterfall" plots. While the field response of strain is relatively uniform (Figure 4a2), the change in lattice tilts is inhomogeneous (Figure 4b2). Specifically, some regions of material remain unchanged in the entire cycle, while other regions show a vigorous response to the field. We labelled the two types of regions as pinned (P) and responsive (R) in Figure 4b2, respectively.

Whether a region is responsive or pinned depends strongly on the local lattice strain under the field-free state. To quantify the cumulative response, we integrated the absolute change of the projected lattice tilt $\sum_i |\boldsymbol{v}(E_{i+1}) - \boldsymbol{v}(E_i)|$ over the entire field cycle (Figure 4e) with the red and blue colours corresponding to the most and least responsive regions, respectively. The absolute value of the strain distribution is also calculated (Figure 4f) and a clear anticorrelation exists between the dynamic field-induced lattice tilts and the static-strain distribution, where regions with the most pronounced change in lattice tilts have the least strain. The observed anticorrelation demonstrates the spatial heterogeneity of the electromechanical response and sheds light on the origin of the relaxor behaviour. The strain maxima or minima have a larger divergence of lattice tilts (Figure 2d) corresponding to the boundaries between polar laminates. These regions serve as pinning points during the field cycle. In contrast, areas with smaller strain are more likely to be inside the laminate with 90° PNDs and are substantially more responsive under electric field. Nanoscopically, inside the polar laminates, the aforementioned cooperative lattice-tilt rotation from the $M_c$ to the orthorhombic phase[22,49] reduces the energy on the 90° domain walls, allowing them to be more responsive under the electric field.

The dramatic difference between intra- and inter-laminate domain walls is striking and highlights the need for spatially resolving the local material properties in the presence of heterogeneity. That not all PND walls respond alike to the electric field requires experimentally and theoretically sampling a sufficiently large field of view containing a statistically significant number of domain walls, which is often either not feasible or a missed opportunity in previous studies. Our findings were made possible by the high spatial resolution, high sensitivity to the material property (lattice deformation in our case), and the large field of view combined with the application of electric field during our experiment.

In summary, *operando* CND studies revealed the existence of hierarchical polar laminates in PMN-0.32PT as well as their role in the field-induced relaxor response. The polar laminates arise from the hierarchical self-organization of twinned monoclinic PNDs separated by 90° domain walls. The boundaries between laminates consist of alternating 90° and 180° domain walls, leading to maxima and minima of strain distribution as well as the divergence of local lattice tilts. The *operando* nano-diffraction studies further established a direct connection between the heterogeneous field response and the detailed structure of the polar laminates. The regions inside the laminates are the most active while the inter-laminate boundaries are spatially pinned due to the presence of the 180° PND walls. The discovery of polar laminates underscores the cooperation between PNDs in facilitating the electromechanical response and sets up a connection between the nanoscale lattice heterogeneity and the macroscopic material properties. The size of the polar laminates governs the population and density of intra- and inter-laminate PND walls and can be tuned sensitively (*e.g.*, via epitaxial strain across the interface). These observations thus provide a guiding principle for the design and optimization of future relaxors[50–52]. Specifically, our work suggests that it is imperative to tailor the macroscopic relaxor behaviour by tuning the position and density of high strain regions in the material. Such local strain modification can be readily achieved via defect engineering (*e.g.*, using ion implantation) or by changing the substrate morphologies. The approach developed herein, including *operando* CND and the associated analysis, is further applicable to a wide range of quantum and functional materials and it is hoped that this work will inspire future studies toward understanding the role of spatial inhomogeneities in a broad class of materials.

**Methods**
**Epitaxial thin-film synthesis:** Pulsed-laser deposition using a KrF excimer laser (248 nm, LPX 300, Coherent) was used to grow 55 nm $0.68Pb(Mg_{1/3}Nb_{2/3})O_3$-$0.32PbTiO_3$ (PMN-0.32PT) / 25 nm $Ba_{0.5}Sr_{0.5}RuO_3$ (BSRO) heterostructures on $SmScO_3$ (SSO) (110) substrates (CrysTec GmbH). The PMN-0.32PT growth was carried out at a heater temperature of 590°C in a dynamic oxygen pressure of 200 mTorr with a laser fluence of 2.0 $J/cm^2$ and a laser repetition rate of 2 Hz from a ceramic target (Praxair) of the same composition with 10% lead excess to compensate for lead loss during growth. The BSRO growth was carried out at temperature of 800°C in a dynamic oxygen pressure of 20 mTorr with a laser fluence of 1.8 $J/cm^2$ and a laser repetition rate of 2 Hz from a ceramic target (Praxair) of the same composition. Following the growth, the samples were cooled to room temperature at 5°C/min. in a static oxygen pressure of 700 Torr. Details of the fabrication process can be found in **Extended Datasets 1.**

**X-ray reciprocal space mapping studies:** Synchrotron X-ray reciprocal space mapping (RSM) studies were collected from Sector 33-ID of the Advanced Photon Source, Argonne National Laboratory. The X-ray diffraction data were obtained using a six-circle diffractometer. The incident X-ray beam has a photon energy of 10 keV and a beam size of 20 $\mu$m (horizontal) × 10 $\mu$m (vertical), slightly smaller than the size of the capacitors being studied (50 $\mu$m in diameter). An incident photon flux of $3\times10^{10}$ photons/second provided enough intensity to observe the diffuse scattering about the structural Bragg peak, which was captured using a Pilatus EIGER 1M area detector. Three rocking scans in sample theta were performed at $(0\ 0\ L)_{pc}$ with L = 1.92, 1.98, and 2.04 to cover the entire diffraction intensities from the PMN-0.32PT, BSRO, and SSO. From the RSM results, the PMN-0.32PT Bragg peak was found to be well separated from the BSRO and SSO Bragg peaks, allowing for a clean diffraction signal for the X-ray nano-diffraction.

**X-ray coherent nano-diffraction:** The X-ray coherent nano-diffraction experiments were carried out at Sector 26-ID of the Advanced Photon Source and Center for Nanoscale Materials (CNM) at Argonne National Laboratory. A schematic of the experimental geometry is shown (Figure 1a). The PMN-0.32PT sample is mounted in a reflective geometry, with sample theta and detector two-theta tuned to access the $(0\ 0\ 2)_{pc}$ Bragg reflection. A monochromatic X-ray beam with an energy of 10 keV was focused on the sample using a Fresnel zone plate (FZP). A beam stop and an order sorting aperture (OSA) were placed before and after the zone plate, respectively, to keep only the first-order component of the focused beam. The focused beam size was 25 nm in diameter with a divergence of 0.5°. A Pilatus EIGER 1M area detector was placed 1.2 meters away from the sample for resolving the lattice distortions from the nano-diffraction pattern.

The diffraction intensity from the convergent X-ray is mainly distributed within an annulus-shaped region on the detector where the parallel fringes arise from the thickness oscillations (inset, Figure 1a). Analysing the distribution of diffraction intensity allows for quantification of the lattice strain $\epsilon_c$ and tilts away from the average atomic structure (*i.e.*, $\Delta\alpha$, and $\Delta\beta$). To illustrate this, an Ewald's construction in a focused beam condition is considered (**Extended Dataset 4**). The focused X-ray beam setup utilizes a cone of parallel beams with different directions of propagation. A fixed area detector placed in a reflective geometry receives diffracted signals from independent parallel beams, which provides a unique opportunity to quantify the lattice distortion in terms of $\epsilon_c$, $\Delta\alpha$, and $\Delta\beta$. Specifically, a *c*-axis strain $\epsilon_c = \Delta c_{pc}/c_{pc}$ (to the zeroth order) leads to a shift of the central diffraction fringe while the annulus stays intact on the detector. In comparison, a lattice tilt $\Delta\beta = \beta - 90°$ leads to a shift of the absolution position of the annulus on the detector as well as a relative shift of the central-diffraction fringe. The absolution and relative shifts are proportional with a ratio of 2:1. The change in $\Delta\alpha = \alpha - 90°$ corresponds to a shift of the donut normal to the scattering plane without relative fringe motion in the scattering plane. By analysing the movement of the annulus and the fringes, we can unambiguously resolve the local distortion of the *c* axis, including the out-of-plane strain and in-plane lattice tilts (**Extended Datasets 5**).

To quantify the lattice tilt from the diffraction pattern in the actual analysis workflow, a tensor of simulated diffraction patterns in a parameter space of $\epsilon_c$ $\Delta\alpha$, and $\Delta\beta$ is precalculated. The similarity between measured and simulated diffraction patterns is estimated through a pixel-to-pixel correlation, and a convolution between measured diffraction pattern and the simulated tensor generates a correlation distribution in the parameter space. Each intensity in such a distribution corresponds to the similarity between the measured diffraction pattern and the simulation with certain lattice distortion. Therefore, the most-matched combination of lattice distortion is identified by finding the maximum correlation intensity in the parameter space (**Extended Datasets 6**).

**Acknowledgements**

The coherent X-ray experiment, data analysis and interpretation at the Argonne National Laboratory were supported by the U.S. Department of Energy, Office of Science, Basic Energy Sciences, Materials Science and Engineering Division. The synthesis of complex oxide films for these driven studies was supported by the U.S. Department of Energy, Office of Science, Office of Basic Energy Sciences, under Award Number DE-SC-0012375. Jie.K. acknowledges the support of the Army/DEVCOM/ARL via the Collaborative for Hierarchical Agile and Responsive Materials (CHARM) under cooperative agreement W911NF-19-2-0119. Jiy.K. acknowledges the support of the Army Research Office under Grant W911NF-21-1-0118. L.W.M. acknowledges additional support from the National Science Foundation under Grant DMR-2102895. This research used resources of the Advanced Photon Source, a U.S. Department of Energy (DOE) Office of Science User Facility operated for the DOE Office of Science by Argonne National Laboratory under Contract No. DE-AC02-06CH11357. Work performed at the Center for Nanoscale Materials, a U.S. Department of Energy Office of Science User Facility, was supported by the U.S. DOE, Office of Basic Energy Sciences, under Contract No. DE-AC02-06CH11357. Y.C. acknowledges helpful discussions with Dr. Daniel Phelan at the Argonne National Laboratory and Prof. Wanzheng Hu of the University of Boston. Y. C. and T.D.F. thanks Jonathan Karsch at the University of Chicago for helping with the PMN-0.32PT device at the early stage of the experiment.

**Additional information**
All data are available in the manuscript or supplementary material.

**Competing financial interests**
The authors declare no competing interests.

**References**

1. Bokov, A. A. & Ye, Z.-G. Recent progress in relaxor ferroelectrics with perovskite structure. *J Mater Sci* **41**, 31–52 (2006).
2. Cartwright, J. H. E. & Mackay, A. L. Beyond crystals: the dialectic of materials and information. *Phil. Trans. R. Soc. A.* **370**, 2807–2822 (2012).
3. Keen, D. A. & Goodwin, A. L. The crystallography of correlated disorder. *Nature* **521**, 303–309 (2015).
4. Billinge, S. J. L. & Levin, I. The Problem with Determining Atomic Structure at the Nanoscale. *Science* **316**, 561–565 (2007).
5. Dagotto, E. Complexity in Strongly Correlated Electronic Systems. *Science* **309**, 257–262 (2005).
6. Choi, S. W., Shrout, R. T. R., Jang, S. J. & Bhalla, A. S. Dielectric and pyroelectric properties in the $Pb(Mg_{1/3}Nb_{2/3})O_3$-$PbTiO_3$ system. *Ferroelectrics* **100**, 29–38 (1989).

7. Cowley, R. A., Gvasaliya, S. N., Lushnikov, S. G., Roessli, B. & Rotaru, G. M. Relaxing with relaxors: a review of relaxor ferroelectrics. *Advances in Physics* **60**, 229–327 (2011).
8. Smolenskii, G. A. & Agranovskaya, A. I. Dielectric polarization of a number of complex compounds. *Phys. Solid State* **1**, 1429 (1959).
9. Cross, L. E. Relaxorferroelectrics: An overview. *Ferroelectrics* **151**, 305–320 (1994).
10. Bokov, A. A. & Ye, Z.-G. DIELECTRIC RELAXATION IN RELAXOR FERROELECTRICS. *J. Adv. Dielect.* **02**, 1241010 (2012).
11. Sun, E. & Cao, W. Relaxor-based ferroelectric single crystals: Growth, domain engineering, characterization and applications. *Progress in Materials Science* **65**, 124–210 (2014).
12. Ye, Z.-G. & Dong, M. Morphotropic domain structures and phase transitions in relaxor-based piezo-/ferroelectric (1−x)Pb(Mg1/3Nb2/3)O3−xPbTiO3 single crystals. *Journal of Applied Physics* **87**, 2312–2319 (2000).
13. Li, F. *et al.* Ultrahigh piezoelectricity in ferroelectric ceramics by design. *Nature Mater* **17**, 349–354 (2018).
14. You, H. & Zhang, Q. M. Diffuse X-Ray Scattering Study of Lead Magnesium Niobate Single Crystals. *Phys. Rev. Lett.* **79**, 3950–3953 (1997).
15. Ye, Z.-G., Noheda, B., Dong, M., Cox, D. & Shirane, G. Monoclinic phase in the relaxor-based piezoelectric/ferroelectric $Pb(Mg_{1/3}Nb_{2/3})O_3 - PbTiO_3$ system. *Phys. Rev. B* **64**, 184114 (2001).
16. Xu, G. *et al.* Ground state of the relaxor ferroelectric $Pb(Zn_{1/3}Nb_{2/3})O_3$. *Phys. Rev. B* **67**, 104102 (2003).
17. Xu, G., Zhong, Z., Hiraka, H. & Shirane, G. Three-dimensional mapping of diffuse scattering in $Pb(Zn_{1/3}Nb_{2/3})O_3 - xPbTiO_3$. *Phys. Rev. B* **70**, 174109 (2004).
18. Xu, G., Zhong, Z., Bing, Y., Ye, Z.-G. & Shirane, G. Electric-field-induced redistribution of polar nano-regions in a relaxor ferroelectric. *Nature Mater* **5**, 134–140 (2006).
19. Matsuura, M. *et al.* Composition dependence of the diffuse scattering in the relaxor ferroelectric compound $(1-x)Pb(Mg_{1/3}Nb_{2/3})O_3 - xPbTiO_3$ ($0 \le x \le 0.40$). *Phys. Rev. B* **74**, 144107 (2006).
20. Xu, G. Probing local polar structures in PZN-*x* PT and PMN-*x* PT relaxor ferroelectrics with neutron and x-ray scattering. *J. Phys.: Conf. Ser.* **320**, 012081 (2011).
21. Krogstad, M. J. *et al.* The relation of local order to material properties in relaxor ferroelectrics. *Nature Mater* **17**, 718–724 (2018).
22. Kim, J. *et al.* Coupled polarization and nanodomain evolution underpins large electromechanical responses in relaxors. *Nat. Phys.* **18**, 1502–1509 (2022).
23. Hirota, K., Ye, Z.-G., Wakimoto, S., Gehring, P. M. & Shirane, G. Neutron diffuse scattering from polar nanoregions in the relaxor $Pb(Mg_{1/3}Nb_{2/3})O_3$. *Phys. Rev. B* **65**, 104105 (2002).
24. Ohwada, K., Hirota, K., Rehrig, P. W., Fujii, Y. & Shirane, G. Neutron diffraction study of field-cooling effects on the relaxor ferroelectric $Pb[(Zn_{1/3}Nb_{2/3})_{0.92}Ti_{0.08}]O_3$. *Phys. Rev. B* **67**, 094111 (2003).
25. La-Orauttapong, D. *et al.* Neutron scattering study of the relaxor ferroelectric $(1-x)Pb(Zn_{1/3}Nb_{2/3})O_3 - xPbTiO_3$. *Phys. Rev. B* **67**, 134110 (2003).
26. Xu, G., Shirane, G., Copley, J. R. D. & Gehring, P. M. Neutron elastic diffuse scattering study of $Pb(Mg_{1/3}Nb_{2/3})O_3$. *Phys. Rev. B* **69**, 064112 (2004).
27. Bai, F. *et al.* X-ray and neutron diffraction investigations of the structural phase transformation sequence under electric field in 0.7Pb(Mg1⁄3Nb2⁄3)-0.3PbTiO3 crystal. *Journal of Applied Physics* **96**, 1620–1627 (2004).
28. Singh, A. K., Pandey, D. & Zaharko, O. Powder neutron diffraction study of phase transitions in and a phase diagram of $(1-x)[Pb(Mg_{1/3}Nb_{2/3})O_3] - xPbTiO_3$. *Phys. Rev. B* **74**, 024101 (2006).
29. Gehring, P. M. NEUTRON DIFFUSE SCATTERING IN LEAD-BASED RELAXOR FERROELECTRICS AND ITS RELATIONSHIP TO THE ULTRA-HIGH PIEZOELECTRICITY. *J. Adv. Dielect.* **02**, 1241005 (2012).
30. Phelan, D. *et al.* Phase diagram of the relaxor ferroelectric $(1-x)Pb(Mg_{1/3}Nb_{2/3})O_3 + xPbTiO_3$ revisited: a neutron powder diffraction study of the relaxor skin effect. *Phase Transitions* **88**, 283–305 (2015).
31. Burton, B., Cockayne, E. & Waghmare, U. Correlations between nanoscale chemical and polar order in relaxor ferroelectrics and the lengthscale for polar nanoregions. *Phys. Rev. B* **72**, 064113 (2005).
32. Xu, G., Wen, J., Stock, C. & Gehring, P. M. Phase instability induced by polar nanoregions in a relaxor ferroelectric system. *Nature Mater* **7**, 562–566 (2008).
33. Manley, M. E. *et al.* Giant electromechanical coupling of relaxor ferroelectrics controlled by polar nanoregion vibrations. *Sci. Adv.* **2**, e1501814 (2016).
34. Li, F. *et al.* The origin of ultrahigh piezoelectricity in relaxor-ferroelectric solid solution crystals. *Nat Commun* **7**, 13807 (2016).

35. Li, F., Zhang, S., Xu, Z. & Chen, L. The Contributions of Polar Nanoregions to the Dielectric and Piezoelectric Responses in Domain-Engineered Relaxor-PbTiO$_3$ Crystals. *Adv. Funct. Mater.* **27**, 1700310 (2017).
36. Takenaka, H., Grinberg, I., Liu, S. & Rappe, A. M. Slush-like polar structures in single-crystal relaxors. *Nature* **546**, 391–395 (2017).
37. Chang, W.-Y. *et al.* Patterned nano-domains in PMN-PT single crystals. *Acta Materialia* **143**, 166–173 (2018).
38. Kumar, A. *et al.* Atomic-resolution electron microscopy of nanoscale local structure in lead-based relaxor ferroelectrics. *Nat. Mater.* **20**, 62–67 (2021).
39. Ye, Z.-G. Crystal chemistry and domain structure of relaxor piezocrystals. *Current Opinion in Solid State and Materials Science* **6**, 35–44 (2002).
40. Shvartsman, V. V., Dkhil, B. & Kholkin, A. L. Mesoscale Domains and Nature of the Relaxor State by Piezoresponse Force Microscopy. *Annu. Rev. Mater. Res.* **43**, 423–449 (2013).
41. Shvartsman, V. V. & Kholkin, A. L. POLAR STRUCTURES OF PbMg$_{1/3}$Nb$_{2/3}$O$_3$-PbTiO$_3$ RELAXORS: PIEZORESPONSE FORCE MICROSCOPY APPROACH. *J. Adv. Dielect.* **02**, 1241003 (2012).
42. Li, F., Zhang, S., Damjanovic, D., Chen, L. & Shrout, T. R. Local Structural Heterogeneity and Electromechanical Responses of Ferroelectrics: Learning from Relaxor Ferroelectrics. *Adv. Funct. Mater.* **28**, 1801504 (2018).
43. Winarski, R. P. *et al.* A hard X-ray nanoprobe beamline for nanoscale microscopy. *J Synchrotron Rad* **19**, 1056–1060 (2012).
44. Kim, J. *et al.* Epitaxial Strain Control of Relaxor Ferroelectric Phase Evolution. *Adv. Mater.* **31**, 1901060 (2019).
45. Ho, J. C., Liu, K. S. & Lin, I. N. Study of ferroelectricity in the PMN-PT system near the morphotropic phase boundary. *J Mater Sci* **28**, 4497–4502 (1993).
46. Chu, Y.-H. *et al.* Nanoscale Domain Control in Multiferroic BiFeO3 Thin Films. *Adv. Mater.* **18**, 2307–2311 (2006).
47. Yan, Y. *et al.* Enhanced temperature stability in ⟨111⟩ textured tetragonal Pb(Mg$_{1/3}$Nb$_{2/3}$)O$_3$-PbTiO$_3$ piezoelectric ceramics. *Journal of Applied Physics* **118**, 104101 (2015).
48. Bednyakov, P. S., Sturman, B. I., Sluka, T., Tagantsev, A. K. & Yudin, P. V. Physics and applications of charged domain walls. *npj Comput Mater* **4**, 65 (2018).
49. Fu, H. & Cohen, R. E. Polarization rotation mechanism for ultrahigh electromechanical response in single-crystal piezoelectrics. *Nature* **403**, 281–283 (2000).
50. Martin, L. W. & Rappe, A. M. Thin-film ferroelectric materials and their applications. *Nat Rev Mater* **2**, 16087 (2016).
51. Martin, L. W., Chu, Y.-H. & Ramesh, R. Advances in the growth and characterization of magnetic, ferroelectric, and multiferroic oxide thin films. *Materials Science and Engineering: R: Reports* **68**, 89–133 (2010).
52. Sharma, P., Moise, T. S., Colombo, L. & Seidel, J. Roadmap for Ferroelectric Domain Wall Nanoelectronics. *Adv Funct Materials* **32**, 2110263 (2022).